\documentclass[conference]{IEEEtran}
\IEEEoverridecommandlockouts
\usepackage{cite}
\usepackage{amsmath,amssymb,amsfonts}
\usepackage{algorithmic}
\usepackage{graphicx}
\usepackage{multirow}  
\usepackage{textcomp}
\usepackage{xcolor}
\def\BibTeX{{\rm B\kern-.05em{\sc i\kern-.025em b}\kern-.08em
    T\kern-.1667em\lower.7ex\hbox{E}\kern-.125emX}}

\newcommand{\ignorethis } [1] {}

\newcommand{\Reals      }     {{\textrm{I\kern-0.18em R}}}

\newcommand{\change     } [1] {\mbox{{\footnotesize $\Delta$} \kern-3pt}#1}

\begin{document}

%\title{Estimating Background Acoustic Features in supervised Manner for Non-intrusive
%Speech Quality Assessment \\
%{\footnotesize }
%\thanks{}
%}

%\title{Embedding Background Noise Information into Learned Representations for Improved Non-Intrusive Speech Quality Assessment \\ 
%{\footnotesize }
%\thanks{}
%}

\title{ A Pre-training Framework that Encodes Noise Information for Speech Quality Assessment\\
\thanks{This work was supported in part by the National Science Foundation under grant IIS-2235228, and in part by the Ohio Supercomputer Center.}}

\author{\IEEEauthorblockN{Subrina Sultana}
\IEEEauthorblockA{\textit{Department of Computer Science and Engineering} \\
\textit{The Ohio State University}\\
Columbus, OH, USA \\
sultana.23@osu.edu}
\and
\IEEEauthorblockN{Donald S. Williamson}
\IEEEauthorblockA{\textit{Department of Computer Science and Engineering} \\
\textit{The Ohio State University}\\
Columbus, OH, USA \\
williamson.413@osu.edu}
}

\maketitle

\begin{abstract}
Self-supervised learning (SSL) has grown in interest within the speech processing community, since it produces representations that are useful for many downstream tasks. SSL uses global and contextual methods to produce robust representations, where SSL even outperforms supervised models. Most self-supervised approaches, however, are limited to embedding information about, i.e., the phonemes, speaker identity, and emotion, into the extracted representations, where they become invariant to background sounds due to contrastive and auto-regressive learning. This is limiting because many downstream tasks leverage noise information to function accurately. Therefore, we propose a pre-training framework that learns information pertaining to background noise in a supervised manner, while jointly embedding speech information using a self-supervised strategy. We experiment with multiple encoders and show that our framework is useful for perceptual speech quality estimation, which relies on background cues. Our results show that the proposed approach improves performance with fewer parameters, in comparison to multiple baselines.
\end{abstract}

\begin{IEEEkeywords}
speech quality, self-supervised learning, representation learning, background noise
\end{IEEEkeywords}

\section{Introduction}
 Speech quality assessment is important to speech enhancement, automatic speech recognition (ASR), and speaker verification, because the quality of speech impacts performance. Subjective quality assessment is the gold standard approach for evaluating speech quality, especially, for applications involving human listeners (e.g., virtual meetings, hearing aids, etc.), though it is infeasible to conduct often as they are time-consuming and costly. 
 
 Many deep learning networks have been developed for estimating objective perceptual quality metrics such as the perceptual evaluation of speech quality (PESQ)\cite{fu2018quality,dong2020attention} and short-time objective intelligibility (STOI) \cite{zezario2020stoi, zhang2021end}, and for estimating subjective metrics, such as the mean opinion score (MOS) \cite{nayem2023attention, reddy2021dnsmos, mittag2021nisqa, liu2022bit}, to model human perception. Non-intrusive MOS prediction, which performs objective assessment without a reference signal, has gained in popularity since it is useful to more applications than intrusive approaches, and it is more closely aligned with true subjective quality assessments than objective metrics such as PESQ and STOI. Additionally, they are faster and more cost-effective than formal subjective evaluations, which is beneficial to speech enhancement and separation. %But this objective metric is not always correlated well with subjective quality assessment.  
A variety of deep-learning approaches have been developed for MOS prediction. For instance, MOSNet \cite{mosnet} uses a convolutional and  bidirectional long short-term memory network to assess the quality of voice conversion, NISQA \cite{mittag2021nisqa} uses a convolutional
neural network with self-attention to assess live-talking and simulated speech, and a pyramid recurrent network is used for speech quality assessment in \cite{dong2020pyramid}. An earlier approach uses a classification-aided network with regression tasks for speech quality assessment \cite{dong2019classification}. These models have achieved success, but some approaches rely on large labeled datasets for better performance, where there may be a scarcity of large target datasets in real-world applications. Furthermore, the utility of these approaches has not been shown on multiple datasets, so they may suffer from the domain shift problem.
 
 Self-supervised learning (SSL) has been beneficial to speech processing, because it can achieve similar performance as supervised models even with a subset of labeled data. SSL generates high-level representations in an unsupervised manner, where the representations are useful for many downstream tasks. Wav2vec 2.0\cite{baevski2020wav2vec} and HuBERT\cite{hsu2021hubert} are popular SSL approaches, and they have been used for phoneme recognition \cite{baevski2020wav2vec}, MOS prediction\cite{cooper2022generalization}, and sensitive word detection \cite{liu2022preventing}. Likewise, Mockingjay\cite{liu2020mockingjay} is a masked transformer approach that uses context to help produce high-level speech representations that are beneficial for phoneme and speaker recognition. In general, these speech representations capture long-range temporal relations and contextual cues based on the contrastive and masking frameworks, which help the models generalize. These SSL models are useful for many speech-related downstream tasks, but unfortunately, their performance on perceptual tasks or auditory scene understanding is not as good. We hypothesize that poor performance occurs because, SSL seeks noise invariance through data augmentations, and it currently does not consider background environmental factors that affect auditory scene understanding and perceptual speech quality, where background sounds are crucial to these tasks. %Rather, extracted speech features from Wav2vec 2.0, HuBERT are useful for phoneme recognition, ASR. 
 %
%Despite, many SSL models have been introduced for speech representations, background acoustic environments like noise, reverberation have not been considered in these SSL models. In the real world, distortions such as reverberation, noise are common scenario for recorded signals. Noise is added at different Signal-To-Noise Ratio (SNR) with clean signals and reverberation can be represented as convolution between a clean signal and a Room Impulse Response (RIR). Perceptual speech quality and intelligibility estimation can represent degrading effects for noise\cite{recommendation2001perceptual} and reverberation\cite{jensen2016algorithm} respectively.
\begin{figure}[htbp]
\centerline{\includegraphics[width=85mm,scale=0.5]{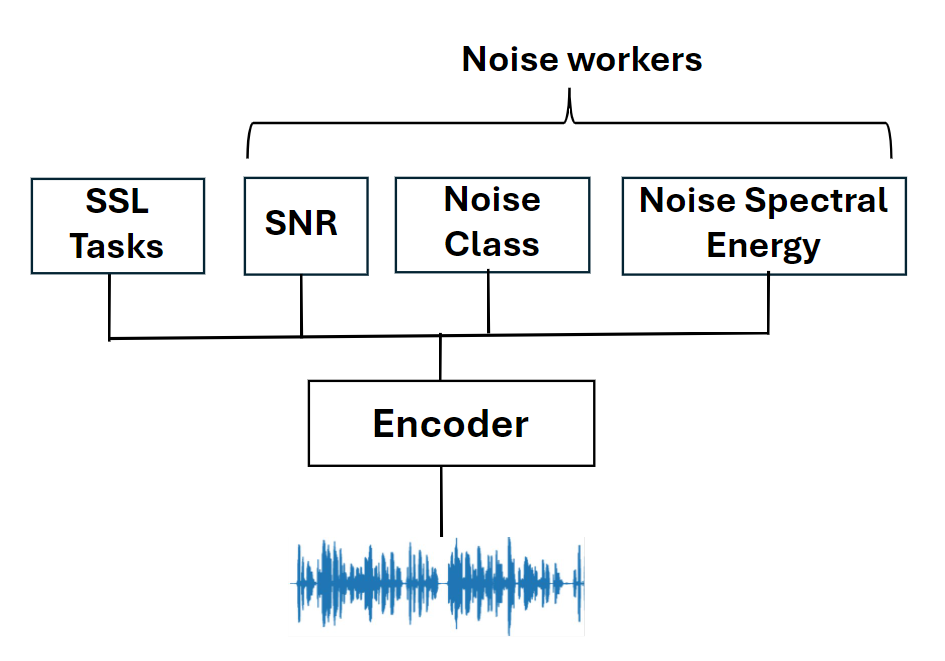}}
\caption{The proposed pre-training framework, where an encoder is pre-trained with proposed noise workers, in addition to the encoder's baseline SSL tasks. The encoder generates latent representations that are used for MOS prediction.}
\label{fig:overview}
\end{figure}

An approach to extract neural embeddings that includes background acoustic properties has been developed \cite{dumpala2024xane}, but it has not been evaluated on relevant downstream tasks. Note that this approach is similar in nature to the approach in \cite{pascual2019learning}, which is a SSL based model that uses an encoder to learn representations that are useful for emotion recognition, ASR, and speaker identification. In both cases, the encoder is optimized with multiple classification and regressive learning objectives, which we denote as workers. %Though PASE encoder can be used for noisy signals during fine-tuning, the SSL model is trained with clean dataset. %t is a multi-task neural network which estimates 14 acoustic parameters incorporating room reverberation and noise. %Accurately estimating these distortions has developed robustness for ASR\cite{li2022non}\cite{parada2015reverberant}. 

 Motivated by \cite{pascual2019learning, dumpala2024xane}, we propose a joint supervised and self-supervised noise-aware pre-training framework that uses an encoder to produce representations that learn information about noise and speech, respectively. Like \cite{pascual2019learning}, an encoder is optimized using multiple self-supervised tasks to account for speech information, but we additional use supervised noise objectives to embed environmental information. Some characteristics such as the noise type, noise spectral pattern, and noise levels are important environmental factors \cite{loizou}. Therefore, we consider three additional classification workers - classifying the category of noise (such as music, human, animal, environment, natural), the signal-to-noise ratio, and the distribution of noise energy in the frequency domain. %We jointly optimize an encoder  so that it learns high-level speech and environmental information. %In addition, from experiments, it can be concluded that pre-trained model containing background acoustic features do well for MOS prediction.

\section{Proposed Model}
An overview of our proposed approach is shown in Fig.~\ref{fig:overview}. An input noisy speech signal is provided to an encoder that produces latent representations. The encoder is optimized using self-supervised and supervised learning tasks, where the former encodes speech information into the representation while the latter encodes noise-level information. This is done during pre-training, where a downstream network later receives the representations and is trained for MOS prediction.

\subsection{Encoders}

To properly evaluate the utility of our approach, we experiment with two different encoders, one is the problem-agnostic speech encoder (PASE) encoder \cite{pascual2019learning} and the other is a masked encoder similar to Mockingjay \cite{liu2020mockingjay} and NISQA \cite{mittag2021nisqa}. We elect to use existing encoders, because our main aim is to determine if noise-related pre-training objectives are beneficial towards the downstream task of MOS prediction. We use the PASE encoder because the latent representation is embedded with low-level and high-level information about the spoken content, and we use a Mockingjay-style encoder since it embeds contextual information using unsupervised learning, where both have performed reasonably well.

\paragraph{Baseline PASE Encoder and Workers}  The PASE encoder consists of a SincNet layer\cite{ravanelli2018speaker}, which performs convolution using parameterized sinc functions. Then seven one-dimensional convolutional blocks are used where each consists of convolution, batch normalization, rectified linear unit (PReLU) activation, and an additional projection layer. % is used to project into final latent representation from 512 features. Convolving a sliding kernel over the speech signal extracts local features at different time frames. 
The encoder outputs a latent representation that is simultaneously provided as input to seven multilayer perceptron workers that perform classification or regressive pre-training tasks. These workers are composed of a single hidden layer (256 units) followed by PReLU activation except for the waveform worker which consists of a waveform decoder followed by a multilayer perceptron. PASE uses four self-supervised regression workers that estimate the: input waveform, log power spectrum (LPS), Mel-frequency cepstral coefficients (MFCC), and prosody, while three self-supervised workers are for binary classification tasks, i.e., local info max (LIM), global info max (GIM), and sequence predicting coding (SPC). The waveform, LPS, MFCC help preserve low-level speech information, and the prosody worker embeds intonation and voicing information.  LIM and GIM help positive samples become closer to a reference anchor compared to negative ones, where positive samples and the anchor come from same speaker. SPC embeds longer context information to maintain sequential ordering. Most regressive workers minimize the mean squared error (MSE) loss, except for the waveform worker that minimizes the L1 loss. The final latent representation has a dimensionality of 100. This matches the reported architecture \cite{pascual2019learning}.  

\paragraph{Masked Encoder and Workers}
 The second encoder is a masking %We experimented with a masked encoder instead of PASE encoder. Our main aim is to test whether accurate reconstruction of original frames can help for downstream task(MOS prediction) through retaining encoder feature representations during fine-tuning. Masked 
 encoder that is composed of multi-layer transformer blocks, and it uses multi-head self-attention. Positional encodings are used to maintain sequential ordering. The Mel-spectrogram is computed from the audio waveform, where we then randomly mask 15\% of the time frames during training, same as\cite{liu2020mockingjay}. For the frames that are masked, 80\% of the time the frame is replaced with zeros, 10\% of the time the frame is replaced with random values, and it is unchanged for the remaining times. We use 3 layers of transformers, 8 attention heads, feed\-forward layer of 1024 nodes, hidden dimension size of 256, and a dropout rate of 0.1. The L1 loss minimizes the error. % between the predicted frames and original frames. %The extracted speech representations contain high-level information and are provided to a downstream task. More details about the approach can be found in \cite{liu2020mockingjay}
  
%The PASE encoder is employed as feature extractor (features do not change dynamically during training fine-tuned model) for downstream task- MOS prediction. In addition, features extracted from self-supervised learning can improve performance for supervised training in low-resource settings. PASE is pre-trained with nine workers and PASE encoder represents low-level and high-level features in addition to noise. 
\begin{figure}[htbp]
\centerline{\includegraphics[width=85mm,scale=0.5]{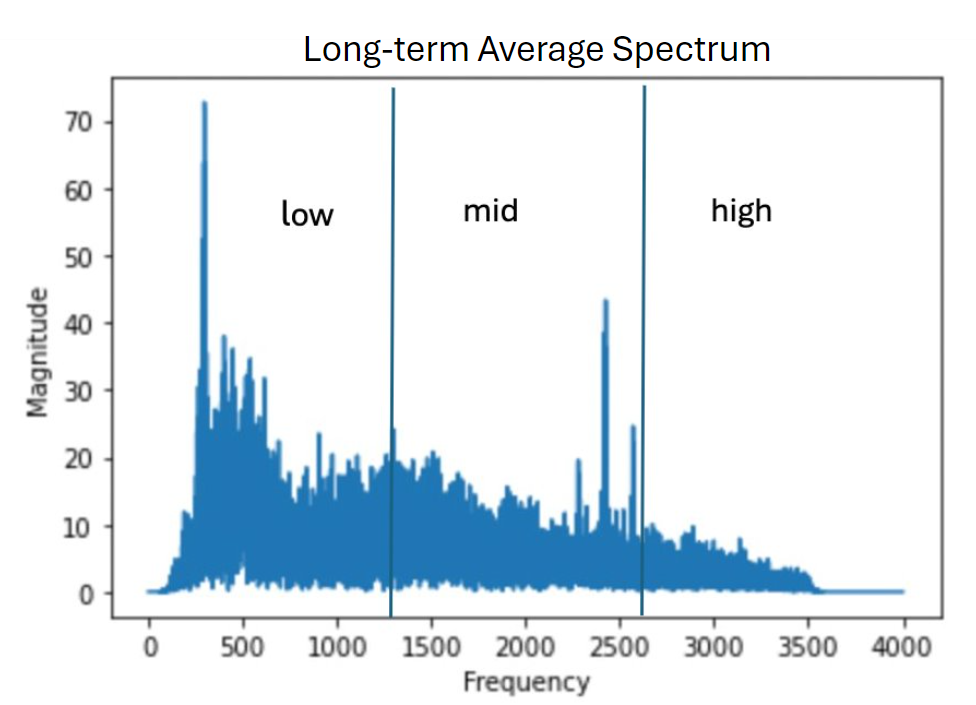}}
\caption{Spectral noise energy distribution of a signal with street noise using a 0dB SNR. The low frequency range has the largest energy.}
\label{fig1}
\end{figure}

\subsection{Proposed Noise Workers}
%The baseline encoders are good at infusing low- and high-level information about the speech, along with contextual information that helps with sequencing. However, these encoders do not learn high-level information about the noise. 
%\vspace{-2.5ex}
We incorporate three supervised classification tasks to embed high-level information about the noise into the encoder's latent representations. We use separate multilayer perceptron workers that each classify the (1) type of signal, (2)  the distribution of the spectral noise energy \cite{loizou}, which also provides information about the type of noise, and the (3) signal-to-noise ratio, which relates to the acoustic environment. %where the classes include: human sounds, source-ambiguous sounds, animal sounds, sounds of things, music sounds, natural sounds, channel, environment and background sounds, based on AudioSet \cite{gemmeke2017audio}. %We separate noisy signals based on these categories. The second classifier looks at the , since the spectral location of the noise provides additional information about the type of noise \cite{loizou}. The third classifier estimates the  %(Shape of Noise, Classify Category of noise). Among them, some of the workers are beneficial for all speech processing and some of them are application dependent. oise shape and category workers help the encoder to retain noise features. This way, regression task of MOS prediction takes advantage of encoder which contains high level speech features. Workers network architecture is kept same as PASE\cite{pascual2019learning}. 

The multi-layer perceptrons each have a single hidden layer with 256 units. The PReLU activation function is used with a cross-entropy loss. he proposed classification workers are described in more detail below. 

\subsubsection{Spectral Energy Distribution}
%Noise can be both stationary (spectral and temporal characteristics do not constantly change over time) and nonstationary (opposite of stationary). In this paper, all noise signals are nonstationary.  Noise has another distinctive feature, 
%The shape of individual spectrums, defined as distribution of noise energy in frequency domain\cite{loizou}. 
We uniformly divide the spectrum of a signal into three regions, denoted as low, mid, and high-frequency regions. %The low region is up to 2667Hz, mid region is from 2668Hz to 5334Hz and high region from 5335Hz to 8kHz by dividing frequency range into 3 equal ranges. 
The region with the largest energy is denoted as the label for the signal and an one hot encoded vector is used. %So, if the energy is largest in the middle frequencies, the label for this signal would be [0,1,0]. % If it is largest at the high frequencies, it would be [0,0,1].
Fig.~\ref{fig1} shows the spectrum of a noisy signal with the largest energy in the low frequency range. In total, there are four classes, where three classes identify low, mid, and high regions for noisy signals, while the last one is reserved for clean signals. %Finally, train the energy distribution worker that classifies the signal according to the label.  

\subsubsection{Category of noise}
We use seven categories of noise based on AudioSet \cite{gemmeke2017audio}: human, source-ambiguous, animal, sounds of things, music, natural, and background sounds.  In total, there are 8 classes where 7 classes identify the noise category, while the last one denotes clean speech. %Finally, train the category of noise worker that classifies the signal according to the label. The audio sample that falls into one of seven categories is labeled with that category using an one hot encoding target.
 %We used seven categories as target labels such as human sounds, source-ambiguous sounds, animal sounds, sounds of things, music sounds, natural sounds, channel, environment and background sounds from AudioSet ontology. 

\subsubsection{Signal to Noise Ratio (SNR)}
%Signal to Noise Ratio(SNR) classification is used to correctly classify the corresponding SNR for each noisy signal. 
Noise is added at SNRs of -5dB, 0dB, 5dB, 10dB, 15dB. The signals are balanced across these SNRs. In total, 6 classes are used where 5 are based on the above SNRs and the last identifies clean speech. %Finally, train the SNR worker that classifies the signal according to the label. 

\subsection{Encoder Pre-training}
The baseline encoder with the additional noise workers is jointly trained using the loss function defined in \eqref{eq},% and gradients calculated from these workers, are averaged out which is defined as total loss. Baseline encoder parameters are updated based on total loss. 
\begin{equation}
L=L_{ssl} - \alpha L_{eng} -\beta L_{noise} -\gamma L_{snr} \label{eq}
\end{equation}
where $L_{ssl}$ denotes the loss for the baseline encoder that uses self-supervised learning, $L_{eng}$, $L_{noise}$, $L_{snr}$ refer to the losses associated with the spectral energy distribution, noise category, and SNR workers, respectively, based on the cross-entropy loss. Here $\alpha$, $\beta$, $\gamma$ values are each 0.1, which was empirically determined. These hyper-parameters (e.g., $\alpha, \beta, \gamma$) balance each worker's contribution during pre-training. %We keep both noisy signals and clean signals during inference to make it more realistic. 
We separately train each encoder for 60 epochs, while keeping the other training parameters the same as in the original papers. %There are unmasked time frames used during inference time to imitate real-world signals\cite{liu2020mockingjay}. 
Note that the workers are only used during pre-training and are later discarded before downstream MOS training.

\subsection{Downstream MOS Training}
The latent representation produced by the encoder is then used for supervised regression-based MOS prediction, where the encoder is held frozen while training the downstream network. For the masked encoder, masking is not performed during this stage. The downstream network consists of one hidden layer of 64 units followed by layer normalization and ReLU activation. Dropout (0.2) is used before the output layer. Range clipping \cite{shu2022non} helps ensure fixed range values for score prediction (e.g., 1 to 5). The network is trained for 1000 epochs, using a batch size of 16, an Adam optimizer with a weight decay of 0.001, a MSE loss, and a learning rate of 0.00012, all of which are empirically determined. 

\section{Experiments}

\subsection{Pre-training Data}
The encoder is pre-trained using clean and noisy signals. Noise signals come from the FSDKaggle2018 dataset of the DCASE Community 2018 Challenge (TASK 2) \cite{Fonseca2017freesound}. %, where these samples are from FreeSound using labels from Google's AudioSet ontology\cite{gemmeke2017audio}. I
In addition, PNL\cite{hu2010tandem} and NoiseX-92\cite{varga1993assessment} are two other sources used during pre-training. In total 11175 noise files are used. The PASE encoder directly works with the audio waveform, while the masked encoder using the Mel-spectrogram. The sampling rate is 16kHz for all signals. The speech signals are from the LibriSpeech dataset \cite{panayotov2015librispeech}. %featuring diverse scenarios and speaker information and only a subset of the whole dataset is used during pre-training and evaluation. 
In total, 23868 signals are used for training and 2652 signals are used for evaluation.

For downstream MOS prediction, we use the Voice Conversion Challenge(VCC) 2018 corpus,a large-scale crowd-sourced listening dataset\cite{lorenzo2018voice}. We use 5400 training samples, 600 evaluation samples, and 500 testing signals. In addition, we use the NISQA dataset for MOS prediction: NISQA\_TRAIN\_SIM (6000 samples) and NISQA\_VAL\_SIM (2500 samples) in which noise signals are taken from the DNS-Challenge dataset\cite{reddy2020interspeech}. We then test with the NISQA\_TEST\_FOR set (240 samples). 
 
 %NISQA\_TRAIN\_SIM and NISQA\_VAL\_SIM carry simulated distortions e.g. bandpass filter, different kinds of codecs, packet-loss and clipping. For adding real background noise, noise signals are taken from DNS-Challenge datasets\cite{reddy2020interspeech}. NISQA\_TEST\_FOR contains English samples and contain live VOIP calls as well as simulated distortions. 

\subsection{Comparison approaches}
We compare our approach to multiple baselines. The PASE baseline pre-trained with a clean dataset \cite{pascual2019learning}. PASE+~\cite{ravanelli2020multi}, which uses 12 workers that are trained with clean speech contaminated by different speech distortions, such as additive noise, reverberation, overlapping speech, clipping, and temporal/frequency masking. The PASE+ model network encourages the encoder to learn distortion-invariant speech representations. Dasheng \cite{dinkel2024scaling}, an SSL approach that encodes speech, music, and environmental information. Wav2vec 2.0 \cite{baevski2020wav2vec} is a self-supervised learning framework that is designed for generating robust audio representations. HuBERT \cite{hsu2021hubert}, which is another self-supervised learning-based network for speech representations. We used all their pre-trained models except for PASE baseline, which we train on our own. During MOS training, the encoder parameters are not updated, since we believe finetuning the encoder will uniformly improve results for all approaches. Hence it would not contribute to our goal of evaluating noise workers for improved MOS prediction.

\begin{table}[htbp]
\caption{An ablation study based on baseline PASE workers and the proposed noise workers, using the VCC dataset. Workers are systematically removed from the specified encoder.}
\begin{center}
\begin{tabular}{|c|c|c|c|c|}
\hline
\textbf{Model} &  \textbf{MSE $\downarrow$}& \textbf{LCC $\uparrow$}& \textbf{SRCC $\uparrow$} \\
\hline
PASE baseline &0.420 &0.642 & 0.621  \\
- waveform & 0.449 &0.615 & 0.574 \\
- LPS&0.426 &0.635 &0.607  \\
- MFCC &0.415 &0.649 &0.627  \\
- Prosody &0.449 & 0.616&0.569  \\
- SPC &0.442 &0.621 & 0.597 \\
- GIM & 0.441&0.617 & 0.588 \\
- LIM & 0.442& 0.619&0.587  \\
\hline
PASE + Noise workers&0.397 & 0.666 & 0.639  \\
- SNR& \textbf{0.386} & \textbf{0.680} & \textbf{0.655}   \\
- Category of noise& 0.423 & 0.646 & 0.618  \\
- Spectral Energy Distribution &0.430& 0.631 & 0.598  \\
- MFCC, SNR& 0.437 & 0.636 & 0.596  \\
%- LPS, MFCC, SNR&0.419 &0.648 & 0.608   \\
\hline
\end{tabular}
\label{tab1}
\end{center}
\end{table}
\section{Results}

\subsection{Ablation Studies: Discarding workers}

%\subsubsection{PASE Encoder} 
Table~\ref{tab1} shows how much the proposed noise workers and baseline workers contribute towards MOS prediction according to MSE, linear correlation coefficient (LCC), and the Spearman’s rank correlation coefficient (SRCC) for the PASE encoder. %For this, we experimented by discarding workers and retrain the encoder with discarding workers. 
In the lower half of the table, we see that removing the SNR noise worker produces the best results amongst all workers that were removed. In fact, without the SNR worker, our approach achieves the best LCC and SRCC values, and a lower MSE value. The performance degrades when the noise category and spectral energy workers are removed individually. In the top half of the table, systematic removal of the PASE baseline workers is evaluated, where the results indicate that most workers contribute to the overall baseline performance, but MFCC contributes least according to all metrics. However, removing the SNR and MFCC workers did not improve performance for the PASE + noise workers approach. %In addition, we discarded two regression workers, LPS and MFCC, since we did another ablation study on PASE baseline workers (individually discarding each one) and observed that LPS, MFCC contributes least to MOS prediction. Furthermore, it is observed that discarding SNR, LPS, MFCC together does not yield a lower MSE value. 

\subsection{Comparison with other SSL models}
 Table~\ref{tab2} shows MOS prediction results using different approaches on the VCC dataset. We observe that PASE with noise workers (without SNR) outperforms the other models, even SSL models with large parameter and dataset sizes, according to all performance measures. It is also worth noting that the addition of the noise workers improve performance over the PASE and Masked Encoder baselines. We surmise that Dasheng performs well because it embeds environmental information. 
%
 %Table~\ref{tab2}, MOS prediction results using the baseline masked encoder (ME) . %We also experimented masked encoder with PASE baseline workers and noise workers. Masked encoder and workers are jointly trained except that additionally L1 loss is considered to update masked encoder parameters rather than using total loss to update encoder parameters during pre-training. 
The masked encoder (ME) with noise workers does not outperform PASE with noise workers. From these experiments, we can conclude that local features extracted by the PASE encoder (e.g., convolutional blocks) are beneficial for MOS prediction rather than generalizing high level contextual information. We make a fair comparison by using almost the same number of training parameters and the same dataset for each encoder.

We also show experimental results on the NISQA dataset in Table~\ref{tab3}. PASE with noise workers (without SNR) also achieves the best performance. %In addition, it is also observed that it achieved better correlation for seen noise dataset. 
 The PASE baseline did not 
%
%\subsection{Evaluation Metrics}
 \begin{table}[htbp!]
\caption{Comparison with other SSL models using the VCC dataset. ME refers to the masked encoder, ``-" indicates that the number of parameters were not mentioned.}
\begin{center}
\begin{tabular}{|c|c|c|c|c|c|}
\hline
\multirow{ 2}{*}{\textbf{Model}} &  \textbf{Emb. } &  \textbf{Total} & \textbf{MSE $\downarrow$}&  \textbf{LCC $\uparrow$} & \textbf{SRCC $\uparrow$}\\
& \textbf{Dim}& \textbf{Param}&  &  &   \\ \hline
PASE+ &256 & - &0.423 &0.655 & 0.622 \\
asheng(Base)&768 & 86M & 0.406&0.667 &0.634  \\
Wav2vec 2.0(Base)&768 & 95M &0.452&0.624 &0.602  \\
uBERT(Base)&768 & 95M & 0.460 & 0.634 & 0.597 \\
\hline
 baseline & 128 & 2.5M  & 0.576 & 0.440 & 0.448 \\
+ Noise workers& 100 & 9.63M &  0.471 & 0.583 & 0.577 \\ %\textbf{(including SNR)} & & & &  & \\ 
\hline
PASE baseline & 100 &12.88M&0.420 &0.642 & 0.621  \\
PASE+Noise workers &100 & 12.94M& \textbf{0.386} &\textbf{0.680} &\textbf{0.655}  \\ 
\hline
\end{tabular}
\label{tab2}
\end{center}
\end{table}
\begin{table}[htbp]
\caption{Performance evaluation using the NISQA dataset}
\begin{center}
\begin{tabular}{|c|c|c|c|}
\hline
\textbf{Model} &  \textbf{MSE $\downarrow$}& \textbf{LCC $\uparrow$}& \textbf{SRCC $\uparrow$} \\
\hline
PASE basleine& 1.132 & 0.516 & 0.494  \\
PASE+Noise workers(Ours)& \textbf{0.448}& \textbf{0.752}& \textbf{0.754}  \\
\hline
Masked Encoder& 1.471 &0.057 &0.074  \\
Masked Encoder+Noise workers& 1.149 &0.453 &0.443   \\
\hline
\end{tabular}
\label{tab3}
\end{center}
\end{table}
work well as it is only trained with a clean dataset, as was done originally. From these experiments, it can be concluded that an encoder containing speech as well as noise information is beneficial for MOS prediction rather than carrying only speech information.

%\subsubsection{Experiment with Masked Encoder}
 %Some of the reasons why masked encoder did not work well - shallow network, comparatively smaller dataset, randomly masking original frames (may contain noise features) and might not reconstruct those time frames properly which have impacts on MOS prediction. 
%From this experiment, it is observed that for masked encoder, it needs lots of training data as well as large network.

\section{Conclusion}
Most SSL models generate embeddings that contain high-level speech information, using autoregressive, contrastive, and masking frameworks. They do not, however, consider background noise which affects speech quality, since a common goal of SSL is producing noise-invariant representations. Our proposed approach generates embeddings that contain information about the speech and background sounds, which are important to perceptual speech quality. From experimental results, we conclude that noise information improves performance for quality assessment, using a simple downstream network. In future work, we will consider other background distortions, environment workers, and we will consider speech intelligibility assessment.

%\section*{Acknowledgment}
%This work was supported in part by the National Science Foundation under grant IIS-2235228, and in part by the Ohio Supercomputer Center.

\bibliographystyle{ieeetr}
\bibliography{ref}

\begin{thebibliography}{10}

\bibitem{fu2018quality}
S.-W. Fu, Y.~Tsao, H.-T. Hwang, and H.-M. Wang, ``Quality-net: An end-to-end non-intrusive speech quality assessment model based on blstm,'' in {\em Interspeech}, pp.~1873--1877, 2018.

\bibitem{dong2020attention}
X.~Dong and D.~S. Williamson, ``An attention enhanced multi-task model for objective speech assessment in real-world environments,'' in {\em IEEE International Conference on Acoustics, Speech and Signal Processing (ICASSP)}, pp.~911--915, IEEE, 2020.

\bibitem{zezario2020stoi}
R.~E. Zezario, S.-W. Fu, C.-S. Fuh, Y.~Tsao, and H.-M. Wang, ``Stoi-net: A deep learning based non-intrusive speech intelligibility assessment model,'' in {\em Asia-Pacific Signal and Information Processing Association Annual Summit and Conference (APSIPA ASC)}, pp.~482--486, IEEE, 2020.

\bibitem{zhang2021end}
Z.~Zhang, P.~Vyas, X.~Dong, and D.~S. Williamson, ``An end-to-end non-intrusive model for subjective and objective real-world speech assessment using a multi-task framework,'' in {\em IEEE International Conference on Acoustics, Speech and Signal Processing (ICASSP)}, pp.~316--320, IEEE, 2021.

\bibitem{nayem2023attention}
K.~M. Nayem and D.~S. Williamson, ``Attention-based speech enhancement using human quality perception modelling,'' {\em IEEE/ACM Transactions on Audio, Speech, and Language Processing}, pp.~250--260, 2023.

\bibitem{reddy2021dnsmos}
C.~K. Reddy, V.~Gopal, and R.~Cutler, ``Dnsmos: A non-intrusive perceptual objective speech quality metric to evaluate noise suppressors,'' in {\em IEEE International Conference on Acoustics, Speech and Signal Processing (ICASSP)}, pp.~6493--6497, IEEE, 2021.

\bibitem{mittag2021nisqa}
G.~Mittag, B.~Naderi, A.~Chehadi, and S.~M{\"o}ller, ``Nisqa: A deep cnn-self-attention model for multidimensional speech quality prediction with crowdsourced datasets,'' in {\em Interspeech}, pp.~2127--2131, 2021.

\bibitem{liu2022bit}
M.~Liu, J.~Wang, L.~Xu, J.~Zhang, S.~Li, and F.~Xiang, ``Bit-mi deep learning-based model to non-intrusive speech quality assessment challenge in online conferencing applications.,'' in {\em INTERSPEECH}, pp.~3288--3292, 2022.

\bibitem{mosnet}
C.-C. Lo, S.-W. Fu, W.-C. Huang, X.~Wang, J.~Yamagishi, Y.~Tsao, and H.-M. Wang, ``Mosnet: Deep learning based objective assessment for voice conversion,'' in {\em Proc. Interspeech}, pp.~1541--1545, 2019.

\bibitem{dong2020pyramid}
X.~Dong and D.~S. Williamson, ``A pyramid recurrent network for predicting crowdsourced speech-quality ratings of real-world signals,'' in {\em INTERSPEECH}, pp.~4631--4635, 2020.

\bibitem{dong2019classification}
X.~Dong and D.~S. Williamson, ``A classification-aided framework for non-intrusive speech quality assessment,'' in {\em IEEE Workshop on Applications of Signal Processing to Audio and Acoustics (WASPAA)}, pp.~100--104, IEEE, 2019.

\bibitem{baevski2020wav2vec}
A.~Baevski, Y.~Zhou, A.~Mohamed, and M.~Auli, ``wav2vec 2.0: A framework for self-supervised learning of speech representations,'' {\em Advances in neural information processing systems}, vol.~33, pp.~12449--12460, 2020.

\bibitem{hsu2021hubert}
W.-N. Hsu, B.~Bolte, Y.-H.~H. Tsai, K.~Lakhotia, R.~Salakhutdinov, and A.~Mohamed, ``Hubert: Self-supervised speech representation learning by masked prediction of hidden units,'' {\em IEEE/ACM transactions on audio, speech, and language processing}, vol.~29, pp.~3451--3460, 2021.

\bibitem{cooper2022generalization}
E.~Cooper, W.-C. Huang, T.~Toda, and J.~Yamagishi, ``Generalization ability of mos prediction networks,'' in {\em IEEE International Conference on Acoustics, Speech and Signal Processing (ICASSP)}, pp.~8442--8446, IEEE, 2022.

\bibitem{liu2022preventing}
Y.~Liu, A.~Kapadia, and D.~S. Williamson, ``Preventing sensitive-word recognition using self-supervised learning to preserve user-privacy for automatic speech recognition.,'' in {\em INTERSPEECH}, pp.~4207--4211, 2022.

\bibitem{liu2020mockingjay}
A.~T. Liu, S.-w. Yang, P.-H. Chi, P.-c. Hsu, and H.-y. Lee, ``Mockingjay: Unsupervised speech representation learning with deep bidirectional transformer encoders,'' in {\em IEEE International Conference on Acoustics, Speech and Signal Processing (ICASSP)}, pp.~6419--6423, IEEE, 2020.

\bibitem{dumpala2024xane}
S.~H. Dumpala, D.~Sharma, C.~S. Sastri, S.~Kruchinin, J.~Fosburgh, and P.~A. Naylor, ``Xane: explainable acoustic neural embeddings,'' in {\em Interspeech}, 2024.

\bibitem{pascual2019learning}
S.~Pascual, M.~Ravanelli, J.~Serra, A.~Bonafonte, and Y.~Bengio, ``Learning problem-agnostic speech representations from multiple self-supervised tasks,'' in {\em Interspeech}, 2019.

\bibitem{loizou}
P.~C. Loizou, {\em Speech Enhancement Theory and Practice, Second Edition}.
\newblock CRC Press, 2013.

\bibitem{ravanelli2018speaker}
M.~Ravanelli and Y.~Bengio, ``Speaker recognition from raw waveform with sincnet,'' in {\em 2018 IEEE spoken language technology workshop (SLT)}, pp.~1021--1028, IEEE, 2018.

\bibitem{gemmeke2017audio}
J.~F. Gemmeke, D.~P. Ellis, D.~Freedman, A.~Jansen, W.~Lawrence, R.~C. Moore, M.~Plakal, and M.~Ritter, ``Audio set: An ontology and human-labeled dataset for audio events,'' in {\em IEEE international conference on acoustics, speech and signal processing (ICASSP)}, pp.~776--780, IEEE, 2017.

\bibitem{shu2022non}
X.~Shu, Y.~Chen, C.~Shang, Y.~Zhao, C.~Zhao, Y.~Zhu, C.~Huang, and Y.~Wang, ``Non-intrusive speech quality assessment with a multi-task learning based subband adaptive attention temporal convolutional neural network.,'' in {\em INTERSPEECH}, pp.~3298--3302, 2022.

\bibitem{Fonseca2017freesound}
E.~Fonseca, J.~Pons, X.~Favory, F.~Font, D.~Bogdanov, A.~Ferraro, S.~Oramas, A.~Porter, and X.~Serra, ``Freesound datasets: a platform for the creation of open audio datasets,'' in {\em Proceedings of the 18th International Society for Music Information Retrieval Conference (ISMIR 2017)}, (Suzhou, China), pp.~486--493, 2017.

\bibitem{hu2010tandem}
G.~Hu and D.~Wang, ``A tandem algorithm for pitch estimation and voiced speech segregation,'' {\em IEEE Transactions on Audio, Speech, and Language Processing}, vol.~18, no.~8, pp.~2067--2079, 2010.

\bibitem{varga1993assessment}
A.~Varga and H.~J. Steeneken, ``Assessment for automatic speech recognition: Ii. noisex-92: A database and an experiment to study the effect of additive noise on speech recognition systems,'' {\em Speech communication}, vol.~12, no.~3, pp.~247--251, 1993.

\bibitem{panayotov2015librispeech}
V.~Panayotov, G.~Chen, D.~Povey, and S.~Khudanpur, ``Librispeech: an asr corpus based on public domain audio books,'' in {\em 2015 IEEE international conference on acoustics, speech and signal processing (ICASSP)}, pp.~5206--5210, IEEE, 2015.

\bibitem{lorenzo2018voice}
J.~Lorenzo-Trueba, J.~Yamagishi, T.~Toda, D.~Saito, F.~Villavicencio, T.~Kinnunen, and Z.~Ling, ``The voice conversion challenge 2018: Promoting development of parallel and nonparallel methods,'' in {\em The Speaker and Language Recognition Workshop}, 2018.

\bibitem{reddy2020interspeech}
C.~K. Reddy, V.~Gopal, R.~Cutler, E.~Beyrami, R.~Cheng, H.~Dubey, S.~Matusevych, R.~Aichner, A.~Aazami, S.~Braun, {\em et~al.}, ``The interspeech 2020 deep noise suppression challenge: Datasets, subjective testing framework, and challenge results,'' in {\em Interspeech}, pp.~2492--2496, 2020.

\bibitem{ravanelli2020multi}
M.~Ravanelli, J.~Zhong, S.~Pascual, P.~Swietojanski, J.~Monteiro, J.~Trmal, and Y.~Bengio, ``Multi-task self-supervised learning for robust speech recognition,'' in {\em IEEE International Conference on Acoustics, Speech and Signal Processing (ICASSP)}, pp.~6989--6993, IEEE, 2020.

\bibitem{dinkel2024scaling}
H.~Dinkel, Z.~Yan, Y.~Wang, J.~Zhang, Y.~Wang, and B.~Wang, ``Scaling up masked audio encoder learning for general audio classification,'' {\em arXiv preprint arXiv:2406.06992}, 2024.

\end{thebibliography}
\vspace{12pt}
\color{red}

\end{document}